# A Robust Dynamic Edge Network Architecture for the Internet-of-Things

Beatriz Lorenzo, Juan García-Rois, Xuanheng Li, Javier González-Castaño, Yuguang Fang

*Abstract*—A massive number of devices are expected to fulfill the missions of sensing, processing and control in cyber-physical Internet-of-Things (IoT) systems with new applications and connectivity requirements. In this context, scarce spectrum resources must accommodate a high traffic volume with stringent requirements of low latency, high reliability and energy efficiency. Conventional centralized network architectures may not be able to fulfill these requirements due to congestion in backhaul links. This article presents a novel design of a robust dynamic edge network architecture (RDNA) for IoT which leverages the latest advances of mobile devices (e.g., their capability to act as access points, storing and computing capabilities) to dynamically harvest unused resources and mitigate network congestion. However, traffic dynamics may compromise the availability of terminal access points and channels and, thus, network connectivity. The proposed design embraces solutions at physical, access, networking, application, and business layers to improve network robustness. The high density of mobile devices provides alternatives for close connectivity which reduces interference and latency, and thus, increases reliability and energy efficiency. Moreover, the computing capabilities of mobile devices project smartness onto the edge which is desirable for autonomous and intelligent decision making. A case study is included to illustrate the performance of RDNA. Potential applications of this architecture in the context of IoT are outlined. Finally, some challenges for future research are presented.

I. INTRODUCTION

The Internet-of-Things (IoT) is a new paradigm that will connect a plethora of physical objects to the Internet and enable them to make intelligent decisions. The underlying technologies of IoT include RFID, sensor networks, pervasive computing, communication technologies, and Internet protocols. In IoT, objects may collaborate and connect to the Internet in a smart way without human intervention to provide new applications. These applications include transportation, manufacturing, healthcare, industrial automation, and emergency handling. The IoT offers great market opportunities and it is expected to have a high economic impact. According to Cisco, there will be 50 billion IoT devices connected to the Internet by 2020 with an expected market volume of $7.1 trillion [1]. The analysis of data trends anticipates that a vast majority of IoT applications will demand high reliability and low latency services from service providers. In this context reliability refers to the capability of

guaranteeing successful message delivery within a given latency bound, and latency refers to the time elapsed since the data is transmitted (e.g., by an object) until it is received by the destination. In Fig. 1, the latency and reliability requirements are shown for the most popular IoT applications with the latency varying between 1ms (ultra-low) and 100ms (low) and the reliability between $1 - 10^{-4}$ (high) and $1 - 10^{-9}$ (ultra-high). Numbers are indicative and may vary for each application area. Mission critical applications such as factory automation require ultra-high reliability and ultra-low latency, while process automation is less demanding. In order to meet these requirements, the evolution of LTE for cellular IoT focuses on extending the battery life of IoT devices and optimizing coverage, capacity and deployment costs with the introduction of eMTC (enhanced Machine-Type Communication) and NB-IoT (Narrow Band-IoT). There can be adopted either on cellular or unlicensed spectrum communications. For further details on communication technologies and standardization efforts for IoT we refer the reader to the survey by Palattella et al. [2]. Despite the existing connectivity solutions, more comprehensive work is needed in the design of new architectures that provision communication and computing resources to successfully support massive IoT deployments. The new architectures should scale operationally and economically with the expansion of IoT and provide smart functionalities for autonomous reasoning among objects.

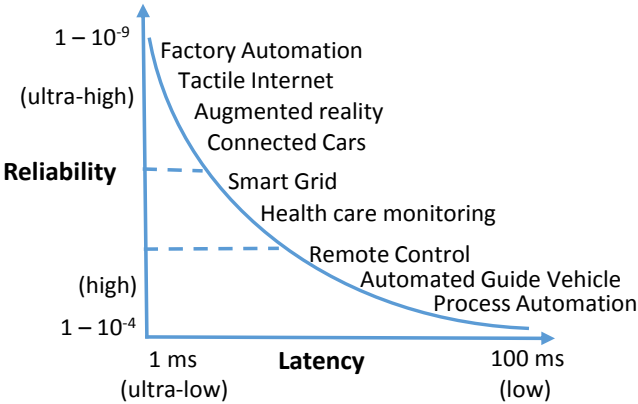

Fig.1. Reliability and latency requirements in IoT.

Several works propose architectures to improve connectivity for specific IoT applications. Castellani et al. have proposed to connect sensors and actuators to the Internet for smart office applications [3]. Similarly, Schleicher et al. [4] have developed an architecture for smart city applications and identified key aspects for its implementation. Wang et al. [16] have presented an energy-efficient architecture for Industrial IoT and proposed a sleep scheduling and wake-up protocol to extend the lifetime of the whole system. Xu et al. [18] have analyzed the integration of IoT with existing networked systems including cloud computing, Internet, smartphone, and industrial networks. With the large scale expansion of IoT, cloud computing based architectures aim to provide complete coverage of processing, computation and storage demands in data centers [5]. However, the centralization of cloud computing and the growing traffic demands of IoT may result in huge bottlenecks degrading network performance. Edge computing architectures could potentially overcome the drawbacks of this approach, moving service provisioning closer to the network edge [6]. Three edge computing architectures have been proposed so far: mobile edge computing, fog computing and cloudlets. Mobile edge computing deploys cloud servers at base stations (BSs) bringing computational capability closer to end-users. The business and technical benefits of mobile edge computing and its integration with IoT are discussed in [19]. The Fog computing edge routers, originally proposed by Cisco for IoT traffic, carry out computing tasks even closer to the users. Sun et al. [20] have proposed a hierarchical fog computing architecture to provide flexible IoT services. They showed that it substantially reduces the traffic load in the core network and the delay between IoT devices and computing resources as compared to traditional IoT architectures. The concept of Cloudlets is an extension of the cloud integrated with Wi-Fi and cellular networks. It is motivated from the fact that it is easy to deploy in coffee shops or office premises for near real-time provisioning.

Regardless of the previous efforts to improve the quality of service in IoT, a holistic approach is needed to improve the robustness of edge computing architectures for IoT to face the dynamic characteristics of wireless network traffic, which result into intermittent connectivity with objects and, consequently, degrade data quality. Furthermore, in the process of moving service provisioning towards the edge, mobile devices keep advancing technologically and getting smarter regarding both their communication and computing capabilities. However, these capabilities have not been fully explored yet in the IoT context.

This article takes a step forward in the design and networking of edge computing wireless architectures for IoT and presents a novel design of a Robust Dynamic Edge Network Architecture (RDNA). The architectural design encompasses solutions at different layers to enhance network robustness against dynamic availability of resources and provide ubiquitous service for IoT. In RDNA, connectivity is provided by users who act as access points for objects and share their storing and computing capabilities, improving scalability, latency, reliability, spectrum efficiency and energy efficiency. Besides, the smart features of mobile devices facilitate autonomous and intelligent decision making at the edge, simplifying control and monitoring of the network.

In the next section, we elaborate the design methodology of RDNA as well as its smart features and applications. Then, we provide some hints for possible business models. Following that, we describe a case study to illustrate its performance in terms of latency, reliability and energy efficiency. Next, we highlight some design challenges and future research directions. Finally, we draw our conclusions.

## II. ROBUST DYNAMIC NETWORK ARCHITECTURE

### A. Preliminaries

Initially Dynamic Network Architectures (DNAs) have been proposed to offload traffic in cellular networks by incentivizing users to share their connectivity and act as access points for neighboring cellular users [7]-[8]. The intermittent availability of these access points renders the architecture dynamic. Shams et al. [8] have developed a framework for topology reconfiguration in DNA and showed that by encoding the problem with a genetic algorithm the optimum topology can track network dynamics and, thus, satisfy users' quality of service (QoS) requirements. Then, Lorenzo et al. [7] have incorporated cognitive capabilities into DNA and showed that underutilized spectrum can be shared temporally and spatially, increasing significantly the network capacity. DNA gains in terms of capacity and revenues make its extension to IoT promising. In particular, the motivation for this extension is threefold. First, the dynamic traffic bursts in IoT can benefit from the additional capacity generated by user participation, anticipating high reuse of network resources. Second, the underlying computing capabilities of user terminals in DNA have not been explored yet and may satisfy the computing needs of IoT traffic. Finally, the smart features of advanced user terminals project smartness onto the edge, which could be exploited to simplify control and monitoring in a DNA with large number of IoT devices. However, to meet the heterogeneous, sometimes exceptionally stringent, requirements of IoT traffic in a dynamic setting a holistic approach to improving network robustness is needed.

*B. Architectural design*

We propose a robust dynamic edge network architecture (RDNA) that leverages the latest advances of wireless devices (e.g., their capability to act as access points, storing and computing capabilities) to provide Internet connectivity and computing capabilities to lightweight IoT objects. The architecture is illustrated in Fig.2 (left). In this architecture, user terminals (phone, PC, tablets) share their connectivities and act as access points for IoT objects for some rewards. We will denote these access points enabled by user terminals as Terminal Access Points (TAPs). In Fig.2, TAPs U1and U3 collect data from objects O1, O2 and O3. The data collected may be used to serve those users or user U2. In a simple business model, an IoT Service Provider (IoTSP) buys data from object owners and organizes the RDNA by incentivizing its users to share their connectivities with IoT objects to meet the demands of an IoT application. RDNA creates an IoT market with opportunities for different providers to cooperate, such as a storage provider negotiating the price to store long term data at user terminals (U1, U3) and distributing that data to other parts of the network (U2), in collaboration with a cloud/fog provider.

The communication and computing capabilities of user terminals are highly heterogeneous and should be managed intelligently to match the heterogeneous demands of IoT traffic. For example, TAPs with wired connectivity (U1, a PC) may serve high priority traffic requiring medium computing processing. Mobile TAPs with wireless connectivity (U3, a phone or tablet) may serve high/medium priority traffic requiring medium/low computing power. This will help to reduce network congestion and preserve dedicated access points or base stations with more powerful computing resources (cloud/fog) for medium/low priority traffic that requires high/medium computing power.

The high density of user terminals brings many opportunities for connectivity and distribution of the collected data throughout the network. Besides, its integration within the cellular infrastructure favors interoperability and seamless addition of heterogeneous IoT objects facilitating scalability and large-scale expansion of RDNA. Furthermore, RDNA improves energy efficiency by utilizing short distance transmissions to reduce interference, and spectrum efficiency by reusing local available channels. In the following we present the design methodology to realize its full potential. Then, we explore the smart features enabled by RDNA, its possible applications and business models.

Since connectivity is provided by mobile user terminals that have limited resources (i.e., battery, CPU, memory) with their own traffic needs, the availability of TAPs and channels is dynamic. This may compromise connectivity and, thus, data quality. In this context, robustness is the property of the network of staying connected which has a direct impact on reliability, latency and energy efficiency, as well as on overall network performance. In the subsequent development, we present design solutions to improve the robustness of RDNA at different levels –physical, access, networking, application, and business. These solutions are illustrated in Fig.2. They may be combined to attain the desired performance.

*Physical Level*

**Data redundancy:** TAPs may aggregate data by collecting it from multiple objects, as shown in Fig.2 (physical level). The degree of data redundancy –repeated data collected for backup and recovery purposes– should be adjusted to satisfy the requirements of the IoT application given the traffic dynamics and to save the energy of objects. An efficient way to control data redundancy and maximize resource utilization is by using pricing schemes. In [9], Luong et al. survey the latest economic models used for data collection in IoT.

*Access Level*

**Cognitive radio capabilities:** interference affects connectivity and degrades quality of communication links. Equipping objects with cognitive capabilities enables them to detect and avoid interference, and opens the possibility of utilizing additionally available communication spectrum. Current regulations of radio spectrum are based on static spectrum allocation policies where spectrum is granted to license holders for long periods of time in large geographic areas. Unfortunately, it has been observed that many allocated spectrum portions are intermittently utilized and so, they may be reused at different times and locations. However, spectrum sensing and spectrum switching processes will consume additional energy and, thus, they may deplete the battery of light-weighted IoT objects. Despite the existing works on energy-efficient and green-energy powered cognitive radio networks, cognitive radio capabilities might not be fully exploited without an appropriate network architecture. In RDNA, TAPs can sense the available spectrum and assist IoT objects without cognitive capabilities so that they could still benefit from cognitive radio technology (e.g., indicating to IoT objects which channel can be used). For more details on cognitive harvesting network architectures, please see [10]. The links established by harvesting available spectrum are referred to as cognitive links. Service providers may also trade available spectrum to meet IoT traffic demands and guarantee required performance levels [7]. In Fig.2 (access level), objects O1 and O2 connect to primary users/TAPs PU1 and PU2 via cognitive links, and O3 also transmits by cognitive links to TAP PU3. The primary users are the licensed holders who have higher priority in the usage of the spectrum while the objects (secondary users) have lower priority. Distant objects may reuse channels to avoid interference.

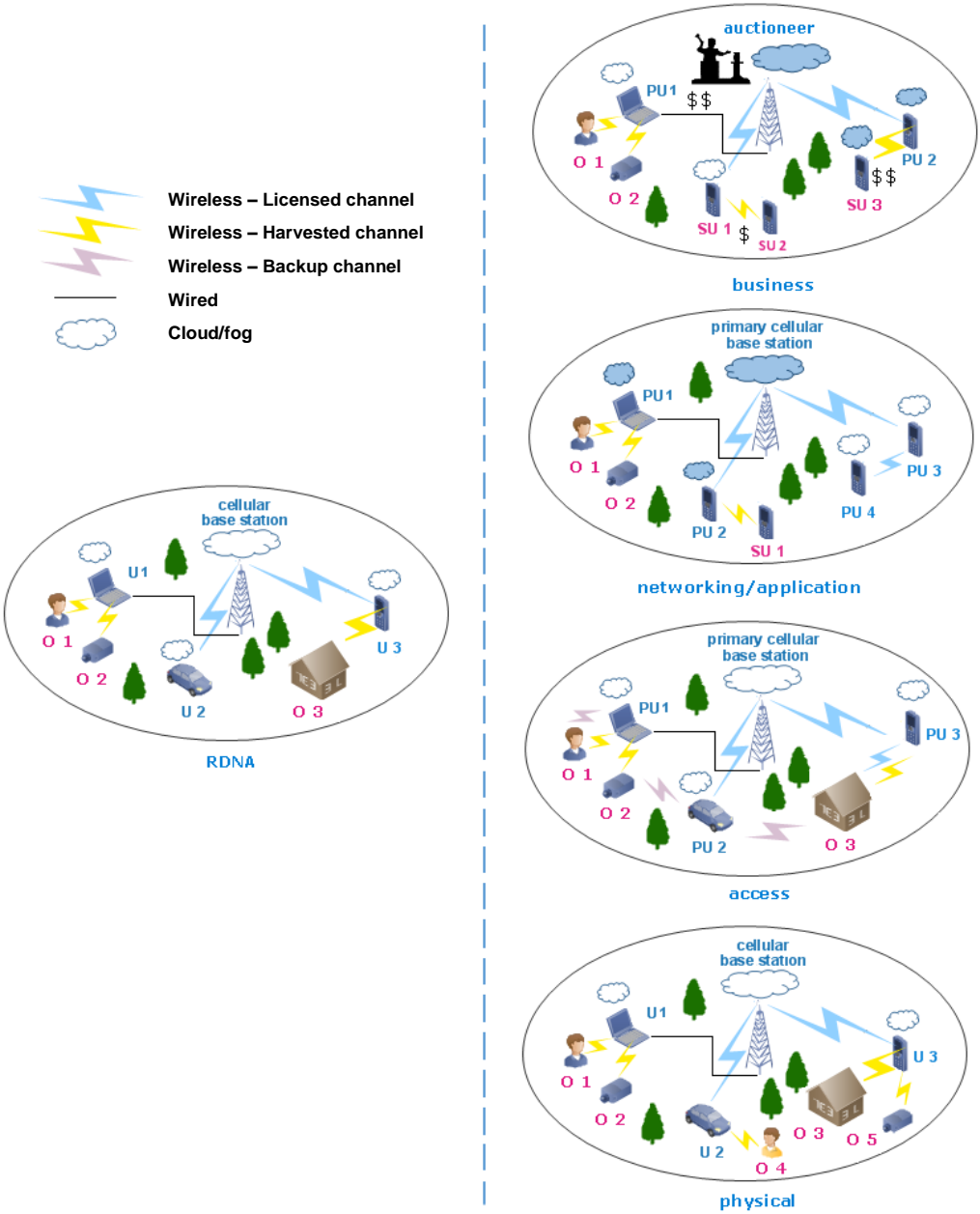

Fig.2. Illustration of RDNA (left) and levels of robustness enhancement of RDNA (right).

**Channel redundancy:** backup channels increase link reliability against traffic dynamics through channel diversity. Backup channels can be licensed or unlicensed to serve high priority or low priority traffic, respectively. A combination of both can be used to achieve different performance and cost tradeoffs. As shown in Fig.2, O1 uses redundant cognitive links to connect to PU1 while O3 has a licensed channel and a backup harvested channel to connect to PU2. A detailed analysis of the benefits of channel redundancy in cognitive networks is given in [11].

**Access point redundancy:** since the availability of TAPs is dynamic, backup access points increase connectivity availability by spatial diversity. TAPs may have wireless or wired connections, resulting in different levels of reliability. This is illustrated in Fig.2 where O2 keeps PU1 and PU2 as backup TAPs and each one has a different type of backhaul connectivity. Object O3 combines the diversity of licensed and unlicensed channels with redundant TAPs PU2 and PU3. A framework to optimize the number of access points to meet quality of service constraints given a fixed cost per access point is elaborated in [8]. Further work on distributed dynamic scheduling schemes is needed to flexibly utilize the available resources [17].

*Networking Level*

**Collaborative data sharing:** subscribers demanding the same data may collaborate by sharing their data using Device-to-Device (D2D) communications and serve as backup for each other's connections. The data can be freely exchanged in a social network or through a content provider that incentivizes users to share their data. The communication can be established by D2D links using licensed or unlicensed channels. For example in Fig.2 (networking level), PU2 and PU3 share their data with SU1 and PU4, respectively. In [7], the benefits of collaborative networking for users and operators are outlined. Epidemic algorithms seem promising for data dissemination in large networks since they are simple to implement and robust against failures.

**Distributed Computing:** mobile terminals may share out their computing capabilities. Since they have heterogeneous capabilities and limited power, highly demanding computing tasks should be decomposed and distributed to different terminals. Distributed computing increases reliability as there is no single point of failure and it is more cost effective than a single high-end computer. In RDNA, a subscriber may request data collected by another user terminal and, on the way, the computing capabilities of the terminals forwarding the data can be exploited. For instance, SU1 may request data collected by PU1 and, thus, PU1 and PU2 may perform related computing tasks. If the computing power is not enough, heavy computing tasks could be delegated to the cloud/fog as illustrated by the colored cloud.

*Application Level*

**Caching:** popular contents can be stored at the network edge to reduce backhaul capacity. Users can share the storage capacity of their terminals temporarily. Unlike collaborative sharing, in caching, users sharing their storage space (PU1, PU2, and PU3) may not be interested in the data to be stored. Cache placement should be optimized to efficiently allocate storage space considering data lifetime, available battery at user terminals and the geographic area where that data is popular. Learning mechanisms may help to decide what content to cache. Dedicated access points and base stations can be used to store long term data and higher data volumes. Bringing content close to end-users reduces latency and traffic load, avoiding duplicate transmissions of the same content across the whole network.

*Business Level*

**Incentives:** Incentive mechanisms are crucial for the adoption and success of RDNA. They should be carefully designed to encourage users to collaborate and share their resources and foster good behaviors. Given the heterogeneity of user terminals and their capabilities, user valuation of its remaining resources should be considered in the design of incentives. Fig.2 (business level) shows the incentives that the IoTSP offers to TAP PU1 to share its connectivity, as well as the revenue earned by SU1 and PU2.

**Multi-provider cooperation:** the IoT market will create business opportunities in RDNA between users and multiple providers. For example, in Fig.2, PU1 provides connectivity to objects O1 and O2 whose data have been requested by secondary users SU2 and SU3. IoTSPs of primary and secondary networks cooperate to serve those users. SU3 will pay more than SU2 for the data since SU3 requests a service that requires additional cloud/fog computing capabilities (colored cloud). The business relationship between the parties involved can be exploited to guarantee service delivery.

*C. Smart features: Automation, Reconfigurability and Intelligence*

User terminals are getting smarter and smarter and can offer substantial storage, communications, control, configuration, measurement and management capabilities at the network edge. Terminals can collect context-aware information regarding traffic demands for the IoTSP to deploy RDNA in congested areas with appropriate robustness. Activating TAPs in such areas contributes to the sustainable deployment of RDNA, since user terminals have limited resources and incentivizing them to cooperate incurs a cost for the IoTSP. By increasing the number of TAPs, the coverage of the IoT as well as the control of the network and physical systems improves. The resulting wider coverage of IoT facilitates remote control of objects and ubiquitous positioning.

Additionally, empowering objects with cognitive capabilities yields high configuration autonomy by dynamic spectrum access, self-adaptation to dynamic scenarios and interference avoidance. A detailed description of how cognitive capabilities contribute to intelligent decision making is given in [12]. Monitoring network conditions (traffic demands, spectrum availability, channel quality) is crucial to reconfigure the network and maintain required reliability and latency levels. The computing power of user devices further broadens reasoning and reconfiguration capabilities of RDNA towards autonomous operations based on the contexts or circumstances. The autonomous control and monitoring of RDNA reduces signaling, which is crucial in presence of a large number of IoT devices.

Providing nodes with information entails excessive signaling and overhead cost. Machine learning schemes such as multi-arm bandits can be used to deal with uncertainty and lack of information to solve resource management problems distributively (e.g., TAPs could make their own decisions about scheduling). Research on intelligent distributed scheduling schemes is left for future work.

The proximity of TAPs accelerates content, services and applications responsiveness from the edge and may allow performing time critical control applications as for example health-care monitoring. Data provided by objects have diverse levels of reliability and trust. Since users value information differently, the intelligence at the network edge can contribute to match user valuations with the expected level of trustworthiness of the data provided. Behavioral game theory is an emerging framework for decision making in the presence of varying levels of intelligence, and hence can be used to study RDNA and the corresponding strategies to boost its performance.

The smart features in RDNA enable highly-improved scalability, adaptability to different situations, and initiation and execution of services with minimal human intervention.

### D. Applications of this architecture

RDNA operates within the existing cellular infrastructure, facilitating device interoperability, and harvesting communication and computing resources network-wide. In the following, we describe some possible applications of this architecture for IoT, which can be used as basis for more specific applications:

**Flexible internetworking:** IoT devices are highly heterogeneous and generate data with different lifespans. RDNA makes internetworking flexible by fulfilling connectivity requirements with TAPs in a dynamic environment without additional infrastructure cost. Therefore, it provides low cost ubiquitous networking towards the Internet-of-Everything (IoE), where objects and humans interconnect seamlessly.

**Computational offloading:** TAPs can perform computing tasks. Choosing among the computing capabilities of TAPs, dedicated access points or base stations will depend on the computational load, tolerated latency and remaining resources at the terminals (e.g., battery, CPU, memory).

**Cloud/fog computing:** fog computing extends cloud computing services to the edge. It improves the efficiency of cloud computing by reducing the amount of data transported to the cloud for processing, analysis and storage. TAPs can be part of the fog paradigm delegating long permanency data to the cloud.

**Collaborative computing:** mobile terminals can combine their computing capabilities to jointly accomplish a common task, increasing their individual computational power.

**Contextual computing:** data collected from objects can be analyzed and combined with context-aware data collected by TAPs themselves. The data collected may include information about user preferences, location and surroundings.

**Transparent computing:** RDNA intelligence enables the network to solve user problems learning from their habits and offer services transparently to end-users.

**Content delivery:** the robust distributed network formed by TAPs allows serving contents to subscribers with high availability and high performance. A content provider can collaborate with an IoTSP to bring contents closer to the edge by storing it in TAPs. In turn, the IoTSP can build a content infrastructure and control network resources for efficient data delivery.

**Mobile big data analytics:** data stored at terminals can be used to extract meaningful data and identify data patterns. This can further increase the intelligence at the edge encouraging autonomy of IoT devices. For instance, data analytics can optimize content delivery by providing insight on the most demanded contents in certain areas of the network.

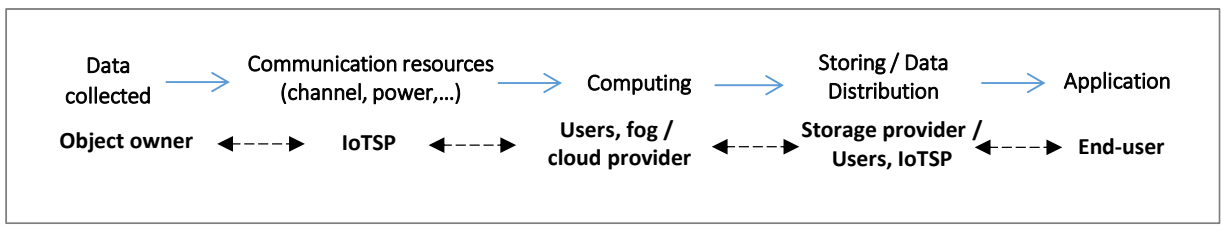

Fig.3. Interaction between different entities in the IoT market.

*E. Business Opportunities*

Internetworking and resource trading at different RDNA levels for the transmission of IoT data create an IoT market with plenty of business opportunities. Fig.3 outlines the interaction among different entities of this market and the resources to trade. IoTSP negotiates with object owners the price for the data to serve its end-users. Since objects have limited capabilities and IoT data is valid for a limited time, IoTSP encourages users to share the connectivity and computing capabilities of their terminals to meet IoT service requirements. If there are not enough terminals nor computing resources available, IoTSP may negotiate with other service providers to access the communication and computing capabilities of their users. The benefits of multi-operator collaboration for data and spectrum trading are described in [7], [13] for different networks. If a data item is requested by end-users that are sparsely distributed throughout the network, IoTSP may negotiate with the fog/cloud provider to distribute the data. Popular data can be cached in TAPs, dedicated access points or base stations depending on user locations. Li et al. [14] discuss pricing and resource allocation in video caching using a game theoretical framework. A more comprehensive survey in [9] summarizes recent economic and pricing models used in IoT for data collection and wireless communications. These models can be extended to RDNA to study topological formation for user terminals collaboration, coverage optimization, computing task allocation, and content distribution. The advantages of economic and pricing models are mainly two-fold. First, they provide a theoretical framework to study revenue generation and thus, analyze benefits and costs. Second, IoT entities have different interests and objectives. Pricing can be used to model the interaction among different entities and encourage them to reach an agreement.

RDNA scales operationally and economically with the proliferation of IoT traffic since its expansion creates more opportunities to collaborate. However, to realize the expected economic impact in IoT, a careful design of economic models is needed to exploit the collaboration opportunities.

## IV. CASE STUDY

We consider a RDNA with objects $\mathcal{I} = \{1, 2, \ldots, n_o\}$ and TAPs $\mathcal{J} = \{1, 2, \ldots, n_{TAP}\}$ that provide Internet connection to those objects and perform pre-processing and data storage. Objects transmit their data using cognitive links in the set $\mathcal{B} = \{1, 2, \ldots, B\}$, which have identical bandwidth equal to 1. RDNA serves $n_s$ end-users who access IoT data through the BS or TAPs. Channel availability varies in time and space as these bands may be occupied by primary users (PUs). We first assume that the IoTSP has no prior knowledge of channel demands of secondary objects (SUs) and later on we utilize the smartness at the edge to monitor such demands. We denote by $a_{ij}^b$ the probability that channel $b$ at link $l_{ij}$ is available for object transmission [11]. Topology formation is based on channel availability and user terminals availability to share their connectivity. The signaling exchange for distributed topology formation has order $\mathcal{O}(n_o n_{TAP} B)$ in the worst case, where there is no prior knowledge on the preferred association. We use an absorbing Markov chain model, where TAPs denote the absorbing states as in [11], and extend it to model the topology evolution in IoT. The latency is obtained as $\tau_j = \tau_o + \tau_p + \tau_a$ where $\tau_o$ is the transmission latency from the object to the TAP, $\tau_p$ is the pre-processing delay and $\tau_a$ is the access delay of each user to RDNA. We assume that the IoTSP has 40 end-users. Fig.4 shows the mean latency versus $n_{TAP}$ for different values of $n_o$. It can be observed that increasing $n_{TAP}$ reduces the latency exponentially. On the other hand, increasing the number of objects $n_o$ increases the latency, especially when $n_{TAP}$ is small. Let us now take advantage of the intelligence provided by TAPs to monitor channel demands of objects. This knowledge is utilized by the IoTSP to reduce the connectivity interruptions by primary user returns and assign channels to the objects which will be available during their transmission period. As shown in Fig.4, smartness at the edge may reduce the latency up to 30% for small $n_{TAP}$. In addition, if users collaborate and share their data using D2D links latency decreases up to 40%. The mean power consumption for transmission, computation and storage is plotted in Fig.5. The transmission power is set to 0.75W and the power consumption for computing and storage varies as in [15]. The power used for channel switching has been neglected. As the transmission distance decreases with $n_{TAP}$, the mean power consumption decreases accordingly.

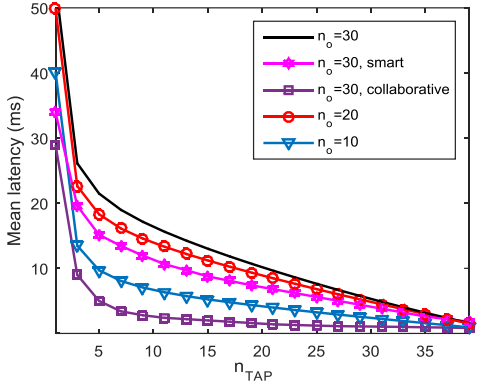 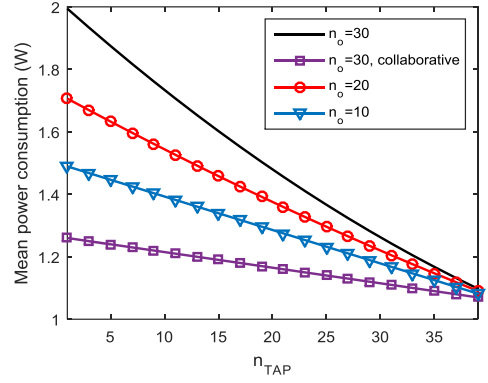

Fig.4. Mean latency vs. number of TAPs.   Fig.5. Mean power consumption vs. number of TAPs.

Let us now determine the redundancy needed to guarantee a level of reliability $\xi_{min}$. We define link reliability $\xi_{ij}$ as the probability that the channel is available (no PU return). We consider that PU arrivals are independent and identically distributed (i.i.d). Hence, link reliability is the same on every channel, $\xi_{ij} = a_{ij}^b$. If a PU returns to the channel currently allocated to a SU, the transmission will be interrupted and the SU will switch to another channel.

*Redundant channels:* As we seek spectrum efficient solutions, the switching interval should be the maximum that satisfies link reliability requirements, so that the number of channels is minimized. Thus, the channel switching time per link per message can be determined as $t_w^* = \mathrm{argmax}_{t_w} \; t_w \cdot (\xi(t_w) - \xi_{min})$, where $t_w^*$ is the maximum duration of the time interval that satisfies $\xi_{min}$. The more restrictive $\xi_{min}$ is, the more often the channel should be switched to avoid PU return. Similarly, for a maximum tolerable latency $\tau_{max}$, the number of backup channels needed can be obtained as $w^* = \mathrm{argmin}_{w \geq w_{min}} (\tau_{max} - \tau(w))^2$, where $\tau(w)$ is the delay when $w$ backup channels are used.

*Redundant terminal access points:* Let us assume that each object selects a set of TAPs on channel $b$ denoted by set $\mathcal{J}_i^b$. Introducing $n_a = |\mathcal{J}_i^b|$ backup TAPs will improve reliability to $\xi_{ij}' = 1 - (1 - \xi_{ij})^{n_a}$. Since energy cost will increase with the additional access points, the objective is to minimize $n_a$ to meet the required reliability level: $(\mathcal{J}_i^b)^* = \mathrm{argmin}_{(\mathcal{J}_i^b)} (\xi'(\mathcal{J}_i^b) - \xi_{min})/n_a$, where $\xi'(\mathcal{J}_i^b)$ denotes the reliability level when the set of backup TAPs $\mathcal{J}_i^b$ is used.

In Fig.6, the number of channels $w$ needed to satisfy $\xi_{min}$ is shown. We assume different ratios of traffic loads in the secondary (objects) and primary network where $\mu_S$ is the service rate of SUs and $\lambda_P$ is the arrival rate of PUs. We can observe that the more imbalanced the network is (i.e., $\mu_S = 6\lambda_P$), the lower is $w$. Besides, increasing the number of backup TAPs significantly reduces $w$. A reliability of $\xi_{min} = 0.999$ can be achieved with 1 to 2 channels and 2 to 3 backup TAPs when $\mu_S = 6\lambda_P$. Besides, by exploiting smartness at the edge to monitor the traffic in the secondary network we can achieve $\xi_{min} = 1$ with only 1 channel. By combining channel and access point redundancy, different reliability requirements can be achieved at a reasonable cost.

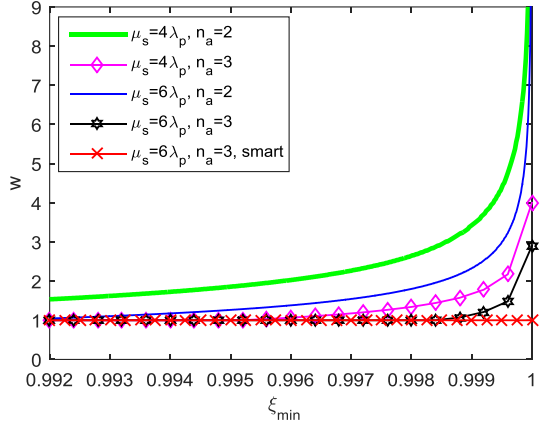

Fig.6. Number of redundant channels vs. required reliability.

## V. DESIGN CHALLENGES AND FUTURE RESEARCH

RDNA moves service provisioning to the edge contributing to the expansion of IoT. It also brings new challenges and promising research directions as highlighted below.

### A. Internet-of-Everything

In the future IoE, human-type communications (HTC) and object-type communications (OTC) will coexist and share limited wireless resources. Each communication type has different needs and capabilities. For example, HTC may require high data rate, while OTC may have stringent low latency requirements. Thus, the main challenge is to develop resource allocation policies that accommodate the heterogeneous nature of IoE traffic with various requirements on service quality.

### B. Security and Privacy

Service providers want to collect as much information as possible regarding user preferences, behavior, and localization, among other metrics, to bring smartness to the edge. This poses a serious challenge for user privacy. Different service providers may need to access content of each other's networks further complicating security concerns. In addition, sharing connectivity and computing capabilities of mobile devices poses extra security design challenges. To protect their privacy, users need to restrict collaboration to other highly reliable and trustworthy users. Likewise, service providers need to ensure the trustworthiness of their subscribers and encourage honest behaviors through proper incentive mechanisms.

*C. Mobility Support*

IoT objects have limited transmission range. Mobility support is important to guarantee their connectivity with mobile TAPs. The smart features of RDNA will help to identify new connectivity options and adapt the connections to the circumstances on the fly to satisfy user demands. Thus, further work is needed in smart topology reconfiguration mechanisms for RDNA in response to mobility.

*D. Business Models*

In the IoT market, entities will change their roles depending on situations and environments. For instance, an IoTSP may act as a data provider or data consumer. A user may buy data and later on sell it to other users. Thus, a sound business model should adapt flexibly to the changing roles of all entities and manage them accordingly.

CONCLUSIONS

This article presents a novel design of a robust dynamic edge network architecture (RDNA) to mitigate the congestion problem in wireless networks, paving the way towards the full realization of IoT. This architectural design leverages the latest technological advances of mobile devices to provide low cost ubiquitous communications and computing. A holistic approach to improving network robustness is developed, which includes solutions at physical, access, networking, application, and business layers. The expected performance in terms of reliability, latency and energy efficiency emphasizes the potential of RDNA for a global IoT architecture. Besides, RDNA brings smart functionalities to the network edge and plenty of business opportunities.